\documentclass{aa}
\usepackage{graphicx}
\usepackage{psfig}

\begin{document}

\def\spose#1{\hbox to 0pt{#1\hss}}
\def\lta{\mathrel{\spose{\lower 3pt\hbox{$\mathchar"218$}}
     \raise 2.0pt\hbox{$\mathchar"13C$}}}
\def\gta{\mathrel{\spose{\lower 3pt\hbox{$\mathchar"218$}}
     \raise 2.0pt\hbox{$\mathchar"13E$}}}
\def\Msun{{\rm M}_\odot}
\def\msun{{\rm M}_\odot}
\def\Rsun{{\rm R}_\odot}
\def\Lsun{{\rm L}_\odot}
\def\half{{1\over2}}
\def\RL{R_{\rm L}}
\def\zs{\zeta_{s}}
\def\zR{\zeta_{\rm R}}
\def\dJJ{{\dot J\over J}}
\def\dMM{{\dot M_2\over M_2}}
\def\tKH{t_{\rm KH}}
\def\eck#1{\left\lbrack #1 \right\rbrack}
\def\rund#1{\left( #1 \right)}
\def\wave#1{\left\lbrace #1 \right\rbrace}
\def\dd{{\rm d}}
\def\new#1{{#1}}

\title{Tracing the power-law component in the energy spectrum of
	black hole candidates as a function of the QPO frequency}

\titlerunning{Tracing the power-law with the QPO frequency}

\author{F. Vignarca\inst{1,2}
	\and
	S. Migliari\inst{1,4}
	\and
        T. Belloni\inst{1}
	\and
	D. Psaltis\inst{3}
	\and
	M. van der Klis\inst{4}
}

\offprints{T. Belloni}

\institute{INAF -- Osservatorio Astronomico di Brera,
	Via E. Bianchi 46, I-23807 Merate (LC), Italy\\
   \and
        Dipartimento di Scienze
	Universit\`a dell'Insubria,
	via Valleggio, I-22100 Como, Italy\\
   \and 
	School of Natural Sciences, Institute for Advanced Study,
	Princeton, NJ 08540, USA\\
   \and
	Astronomical Institute ``Anton Pannekoek" 
	University of Amsterdam and Center for High-Energy Astrophysics,\\ 
        Kruislaan 403, NL 1098 SJ Amsterdam, Netherlands. 
}

\date{Received ???; accepted ???}

\abstract
{
We investigated the relation between the centroid frequency of the
quasi-periodic oscillation observed in the power density spectra
of a sample of galactic black-hole candidates
with the power-law photon index obtained from spectral fits.
Our aim is to avoid
inner accretion disk radius determination directly from spectral
fits, given the uncertainties of the absolute values obtained in 
that way, but to base our analysis on the likely association of 
QPO frequency to a characteristic radius.
We used archival RXTE data of GRS 1915+105 and published parameters
for GRO 1655-40, XTE J1550-564, XTE J1748-288 and 4U 1630-47.
While for low values of the QPO frequency, the two parameters are
clearly correlated for each source, there is evidence for a turnoff
in the correlation above a characteristic frequency, different for 
different sources. We discuss the possible nature of this turnoff. 
} 

\maketitle


\section{Introduction}

In 1992, the WATCH all-sky monitor onboard GRANAT discovered the X-ray source
GRS~1915+105 as a bright transient (Castro-Tirado et al. 1992). Mirabel \&
Rodr\'\i{}guez (1994) reported the observation of radio relativistic
expansions interpreted in terms of collimated emissions of matter, which made
GRS~1915+105 the first galactic object with superluminal jets, for which the
term ``microquasar'' was coined. This source
was originally supposed to host a black hole because 
of its high X-ray luminosity, and for its similarity with another superluminal
source, 
GRO~J1655-40, for which it was possible to obtain a dynamical estimate of the
mass (M=6.3$\pm$0.5M$_{\odot}$) of the compact object (Bailyn et al. 1995;
Greene et al. 2001). No optical
counterpart of GRS~1915+105 was found because of the high Galactic
extinction and only an IR counterpart was detected (Mirabel et al. 1994).
Recently, an orbital period of 33.5 days was discovered with IR spectroscopic
observations, leading to a \new{dynamical}
estimate of 14$\pm$4 M$_\odot$ for the compact
object (Greiner et al. 2001, \new{see also Borozdin et al. 1999}). 
The numerous observations of GRS~1915+105 made with the instruments onboard
the {\it Rossi X-ray Timing Explorer} (RXTE) satellite show an extremely
complex 
variability in the X-ray band, which Belloni at al. (1997a, 1997b) interpreted
as an oscillation of the inner region of the accretion disk caused by a
thermal-viscous instability. Belloni et al. (2000) introduced a
model-independent classification of the different types of X-ray variability
in three basic spectral states.
Similarly to more conventional Black Hole Candidates (BHC), 
the energy spectrum of GRS~1915+105 can
be fitted with the superposition of a soft thermal component, modeled as a
disk-blackbody (Mitsuda et al. 1984), plus a harder component, modeled as a
power law  with a high energy cutoff (Tanaka \& Lewin
1995). The state transition can be attributed to changes in the relative
contributions of these variable components, again similarly to what observed
in other BHCs (Belloni 1998).

Detailed timing analysis of the RXTE data (Morgan et al. 1997)
revealed 
three types of Quasi Periodic Oscillations (QPO): a 1-10~Hz QPO with variable
centroid frequency similar to that observed in other BHCs (see van der Klis
1995), two quasi-stable QPOs with centroid frequencies  $\sim$40 and
$\sim$67~Hz (Morgan et al. 1997; Belloni et al. 2001; 
Strohmayer et al. 2001a), and
variable low frequency QPOs ($10^{-3}-1$~Hz), which can be identified with
transitions between the states of Belloni et al. (2000). 
The presence of the 1-10~Hz is always associated with a spectral state C
(Belloni et al. 2000). 
The 1-10~Hz QPO centroid frequency is well correlated with both disk and
power-law spectral components (Muno et al. 1999,
Reig et al. 2000, Rodriguez et al. 2002) and with the X-ray flux (Chen 
et al. 1997; Markwardt et al. 1999). 

Several theoretical
models associate the origin of this QPO with the innermost region of the disk
(Miller et al. 1998; Nowak \& Wagoner 1991; Titarchuk et al. 1998; Stella \&
Vietri 1998; Stella et al. 1999; 
Varni\`ere et al. 2002, Psaltis \& Norman 2002). 
Di Matteo \& Psaltis (1999) made use of the fact that the fastest variability
timescale at any radius around a compact object is the Keplerian orbital
frequency to obtain upper limits on the radius at which the 1-10 Hz QPO is
produced. As it is apparent in their work, there is a general positive
correlation, considering different systems, between the photon index of the
power-law component in the energy spectrum and the centroid frequency of the
QPO. Sobczak et al. (2000a) in
their analysis of RXTE data of two transient BHCs, XTE~J1550-564 and
GRO~J1655-40, examined the correlation between the spectral parameters,
inner disk radius and power-law photon index, with the QPO centroid
frequency. Surprisingly, they find correlations of opposite sign for the two
sources.
\new{Rodriguez et al. (2002) analyzed the correlation between QPO centroid
frequency and inner disk radius (determined from X-ray spectral fits).}
\begin{table*}[!t]
   $$		
   \begin{array}{clccccccccc}
    \hline
\\     
{\rm Obs. N}&{\rm ~~Observation\, ID}&{\rm MJD}       &{\rm PCA (c/s)}  & {\rm
\Delta T} & \nu_{QPO} & Tim. Res. &  \Gamma  & N_{\rm H} & E_{c} &
\chi_r^{2}\\
	\\
     \hline
&&&&&&&&&&\\
1  & {\rm 10408-01-23-00}a & 50278.4921  &9735  &3072 & 3.491\pm0.005 &
1/512 & 2.62 \pm0.01 & 6.7 & 24.7 & 1.7\\
2  & {\rm 10408-01-23-00}b & 50278.5585  &9710  &3540 & 3.617\pm0.004 &
1/512 & 2.66 \pm0.01 & 6.9 & 25.4 & 1.7\\
3  & {\rm 10408-01-23-00}c & 50278.6266  &10458 &3401 & 4.199\pm0.009 &
1/512 & 2.76 \pm0.01 & 7.1 & 28.0 & 1.7\\
4  & {\rm 10408-01-24-00}a & 50280.1702  &8924  &2497 & 2.228\pm0.008 &
1/512 & 2.38 \pm0.04 & 6.4 & 25.1 & 1.7\\
5  & {\rm 10408-01-24-00}b & 50280.2266  &8901  &3479 & 2.293\pm0.007 &
1/512 & 2.41 \pm0.04 & 6.3 & 23.0 & 1.6\\
6  & {\rm 10408-01-24-00}c & 50280.2933  &8956  &2998 & 2.534\pm0.008 &
1/512 & 2.46 \pm0.04 & 6.6 & 23.0 & 1.4\\
7  & {\rm 10408-01-27-00}a & 50290.5774  &7887  &2421 & 0.642\pm0.004 &
1/512 & 1.88 \pm0.04 & 6.9 & 16.2 & 1.3\\
8  & {\rm 10408-01-27-00}b & 50290.6322  &7926  &3480 & 0.628\pm0.004 &
1/512 & 1.82 \pm0.01 & 6.7 & 15.1 & 1.2\\
9  & {\rm 20402-01-49-01}  & 50730.3949  &8198  &3463 & 2.637\pm0.005 &
1/512 & 2.45 \pm0.01 & 5.4 & 29.7 & 1.8\\
10 & {\rm 20402-01-50-00}  & 50735.5474  &6579  &1962 & 0.835\pm0.010 &
1/128 & 1.92 \pm0.02 & 5.7 & 21.5 & 1.9\\
11 & {\rm 20402-01-50-01}a & 50737.4047  &6608  &3007 & 1.014\pm0.006 &
1/128 & 1.98 \pm0.01 & 5.7 & 22.1 & 1.6\\  
12 & {\rm 20402-01-50-01}b & 50737.4766  &6373  &2620 & 1.078\pm0.005 &
1/128 & 2.00 \pm0.01 & 5.5 & 20.0 & 1.6\\
13 & {\rm 20402-01-52-00}a & 50746.5509  &6657  &2207 & 1.405\pm0.007 &
1/128 & 2.09 \pm0.04 & 5.5 & 25.0 & 1.8\\
14 & {\rm 20402-01-52-00}b & 50746.6190  &6668  &2079 & 1.468\pm0.006 &
1/128 & 2.10 \pm0.04 & 5.5 & 25.1 & 1.4\\
15 & {\rm 20402-01-52-00}c & 50746.6898  &6768  &1725 & 1.609\pm0.007 &
1/128 & 2.16 \pm0.04 & 5.6 & 26.9 & 1.7\\
&&&&&&&&&&\\  
   \hline
   \end{array}
   $$
   \caption[]{List of plateau observations analyzed and results for timing and 
spectral parameters. Letters a, b, c correspond to different orbits in the
same observation. The fifth column refers to the state-C length (in seconds) 
that we
selected for our study. QPO frequencies are in Hertz, N$_{\rm H}\times
10^{22}$cm$^{-2}$. Typical errors are: for N$_{\rm H}\pm$0.1 and for
$E_{c}\pm$0.45}  
       \label{tabobplat}
\end{table*}

In this
paper, we concentrate on the correlation between power-law photon index and
QPO centroid frequency. We present the results of timing and spectral analysis 
of a number of RXTE observations of GRS~1915+105, focusing on this
correlation. We also consider published data relative to four other systems:
GRO~J1655-40 (Remillard et al. 1999a; Sobczak et al. 2000a), XTE~J1550-564
(Sobczak et al. 2000a,b; Homan et al. 2001), XTE~J1748-288 (Revnivtsev et
al. 2000), 4U~1630-47 (Trudolyubov at al. 2001; Tomsick \& Kaaret 2000). 
We then compare the results in term of possible theoretical models. 


\section{Data analysis}

Our study of GRS~1915+105 is divided in two parts, according to the different
classes of observations that we considered. 
In the first part, we analyze the so-called {\it
plateau} observations (Fender et al. 1999; Reig at al. 2000), which
consist in long intervals of persistent state C (class $\chi$ in Belloni et
al. 2000). 
The second part is dedicated to
observations that show large variability in the X-ray flux and at least one 
state-C interval. 
Again following the classification of Belloni et al. (2000), we
selected observations corresponding to classes $\alpha$, $\beta$, and $\nu$. 

\subsection{Plateau observations}

{\it Plateau} intervals show persistent QPOs 
that are very easily detectable in the
Power Density Spectrum (PDS). 
We based our choice of appropriate  observations upon the
extensive analysis of Reig et al. (2000), which allowed us to identify 
the larger possible range of centroid frequencies of the QPO. 
We selected 7 RXTE/PCA observations  of GRS 1915+105 
(for a total of 15 satellite orbits), corresponding to class $\chi$ in the
classification of Belloni et al. (2000). The observation log is
reported in Table~\ref{tabobplat}.

\begin{figure*}[t]
\centering
\begin{tabular}{cc}
\psfig{figure=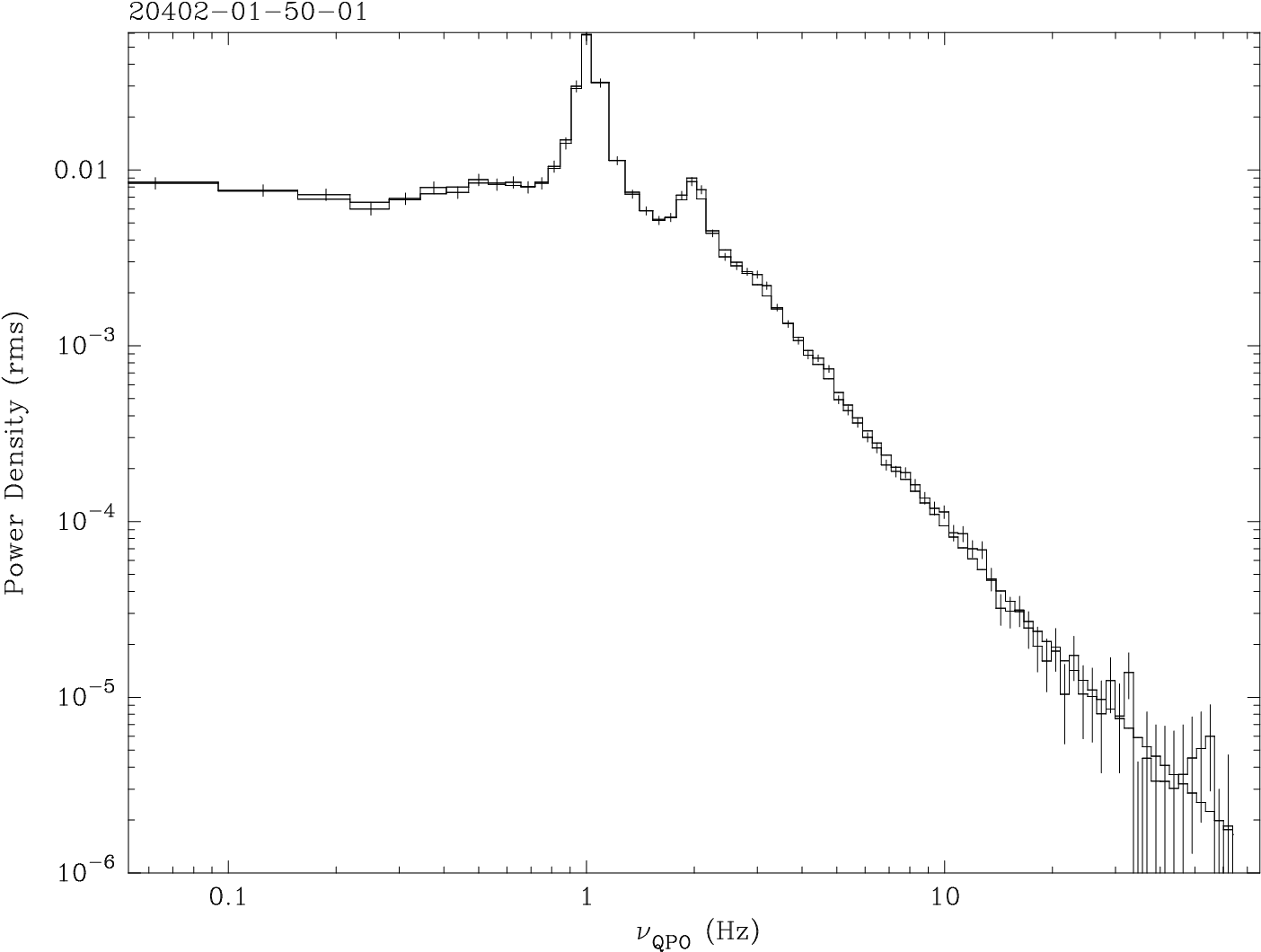,width=6cm,angle=-90,height=4cm}
&
\psfig{figure=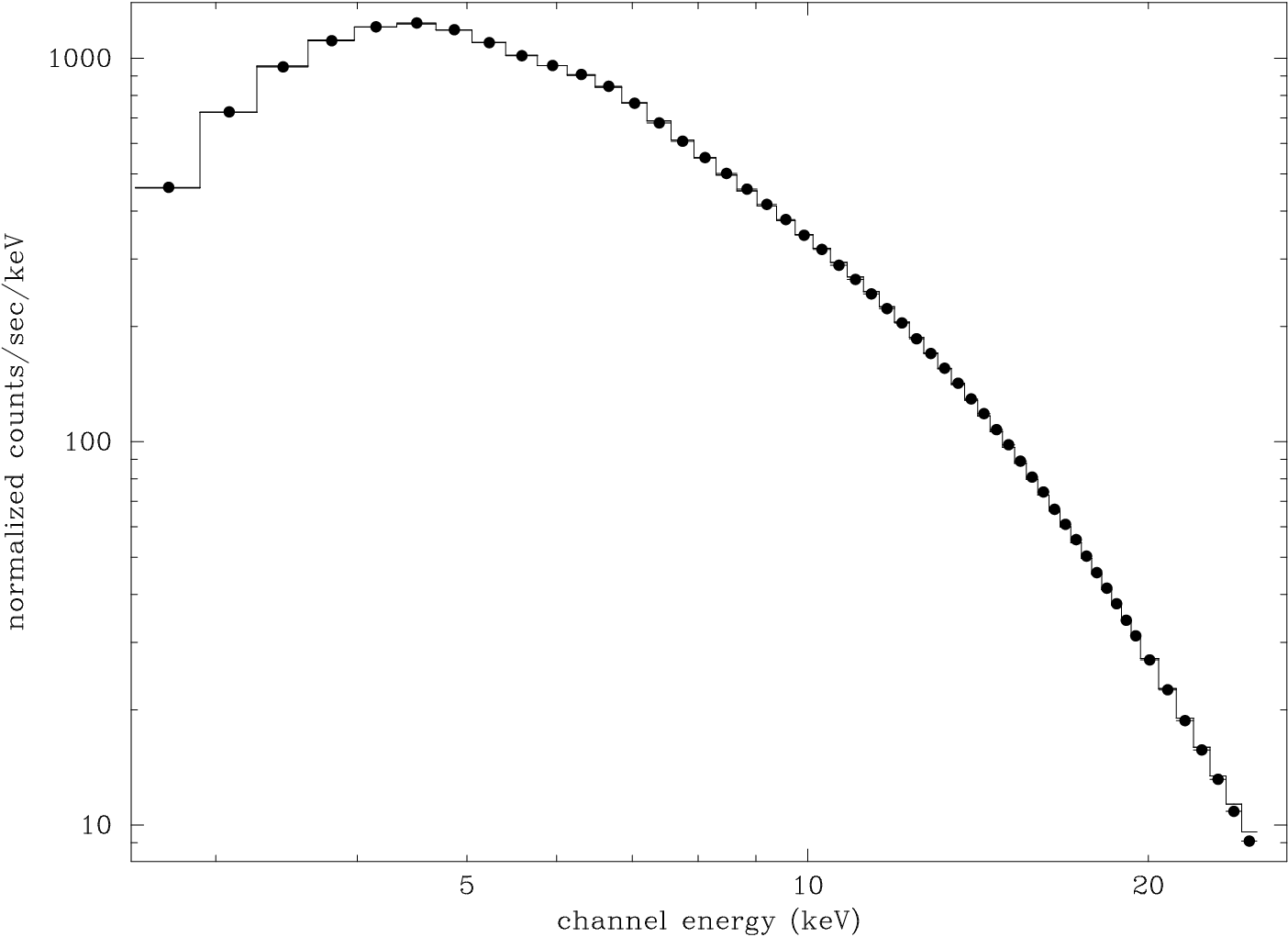,width=6cm,angle=-90,height=4cm}
\\
\psfig{figure=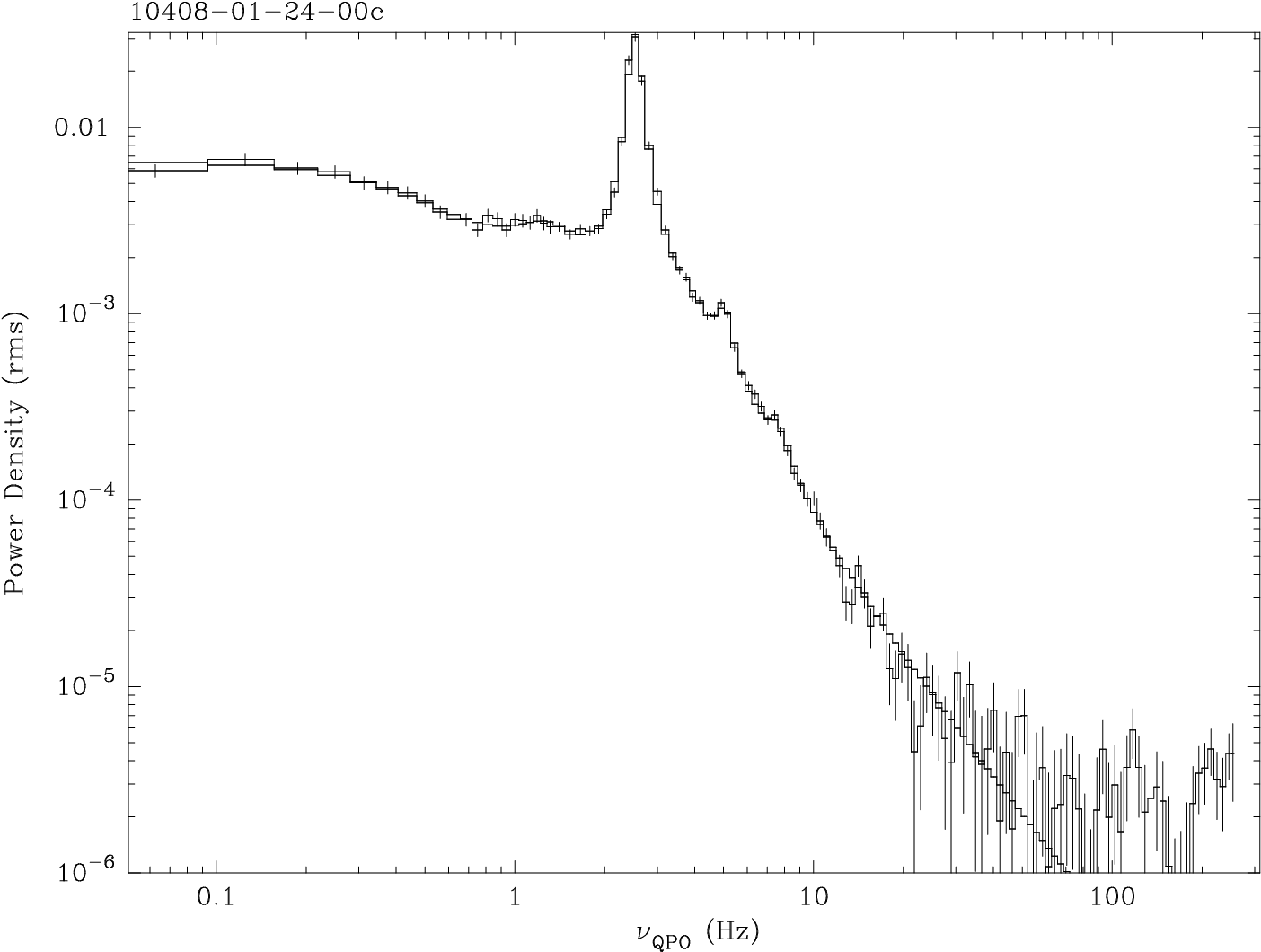,width=6cm,angle=-90,height=4cm}
&
\psfig{figure=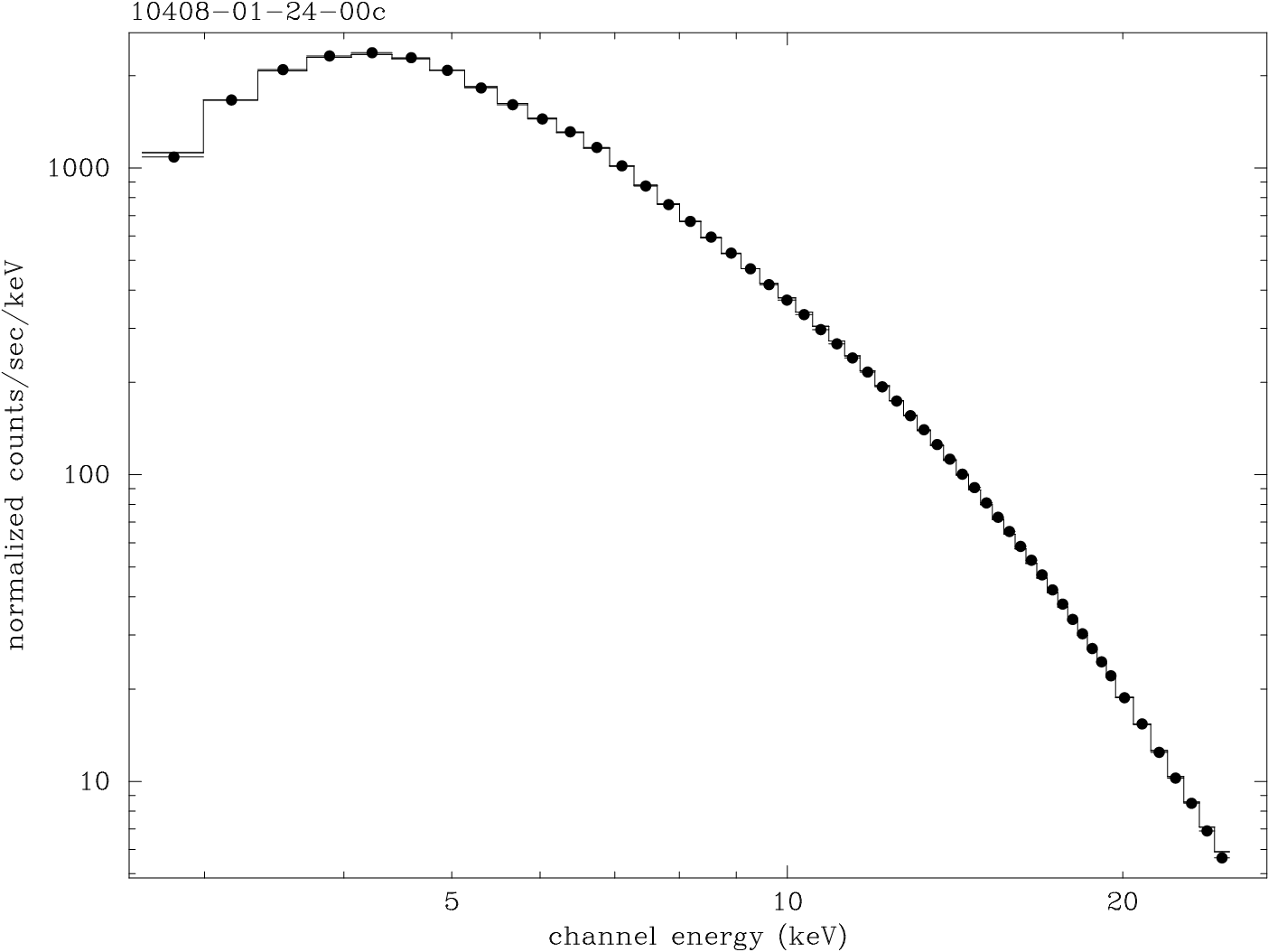,width=6cm,angle=-90,height=4cm}
\\
\psfig{figure=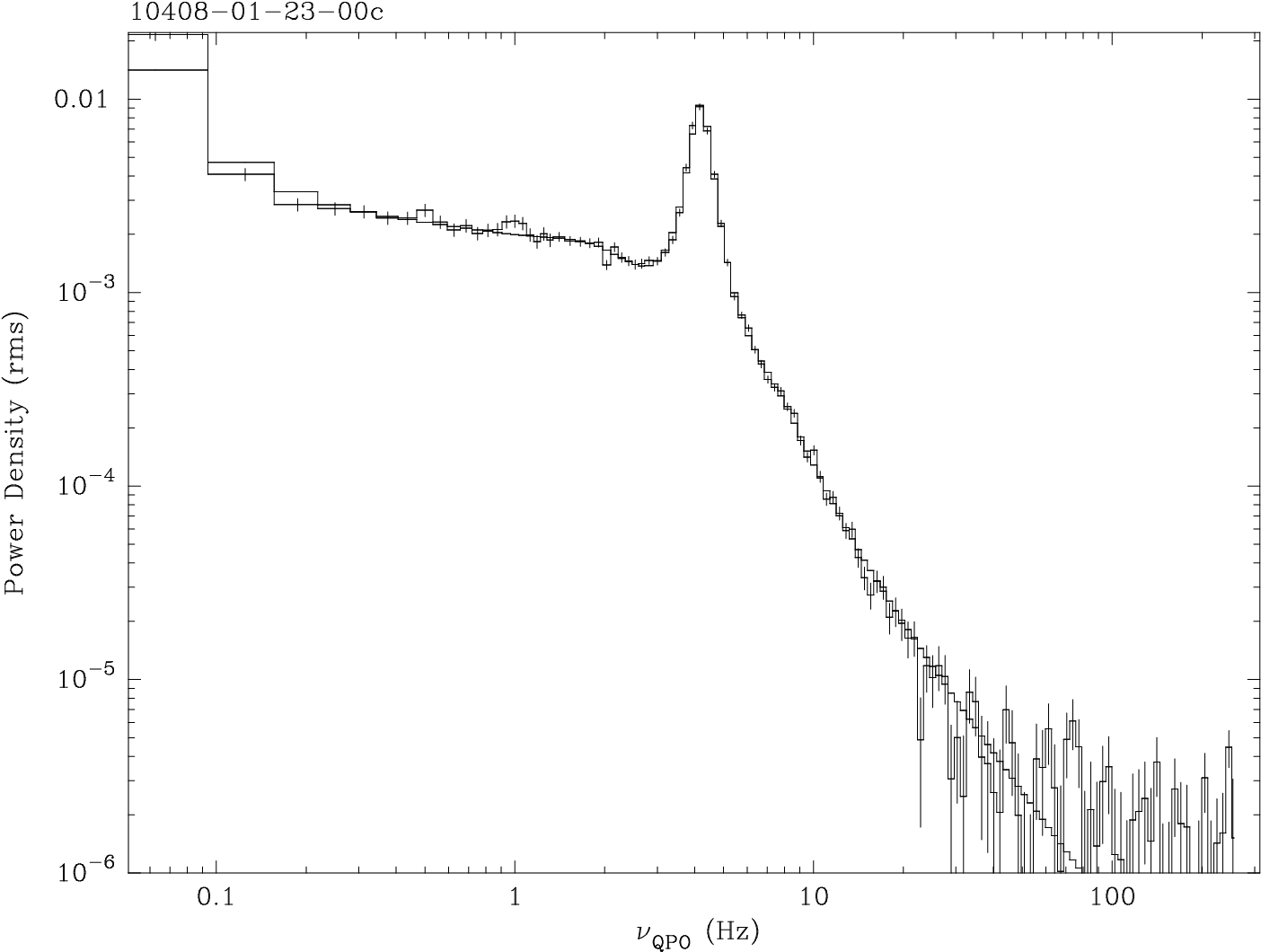,width=6cm,angle=-90,height=4cm}
&
\psfig{figure=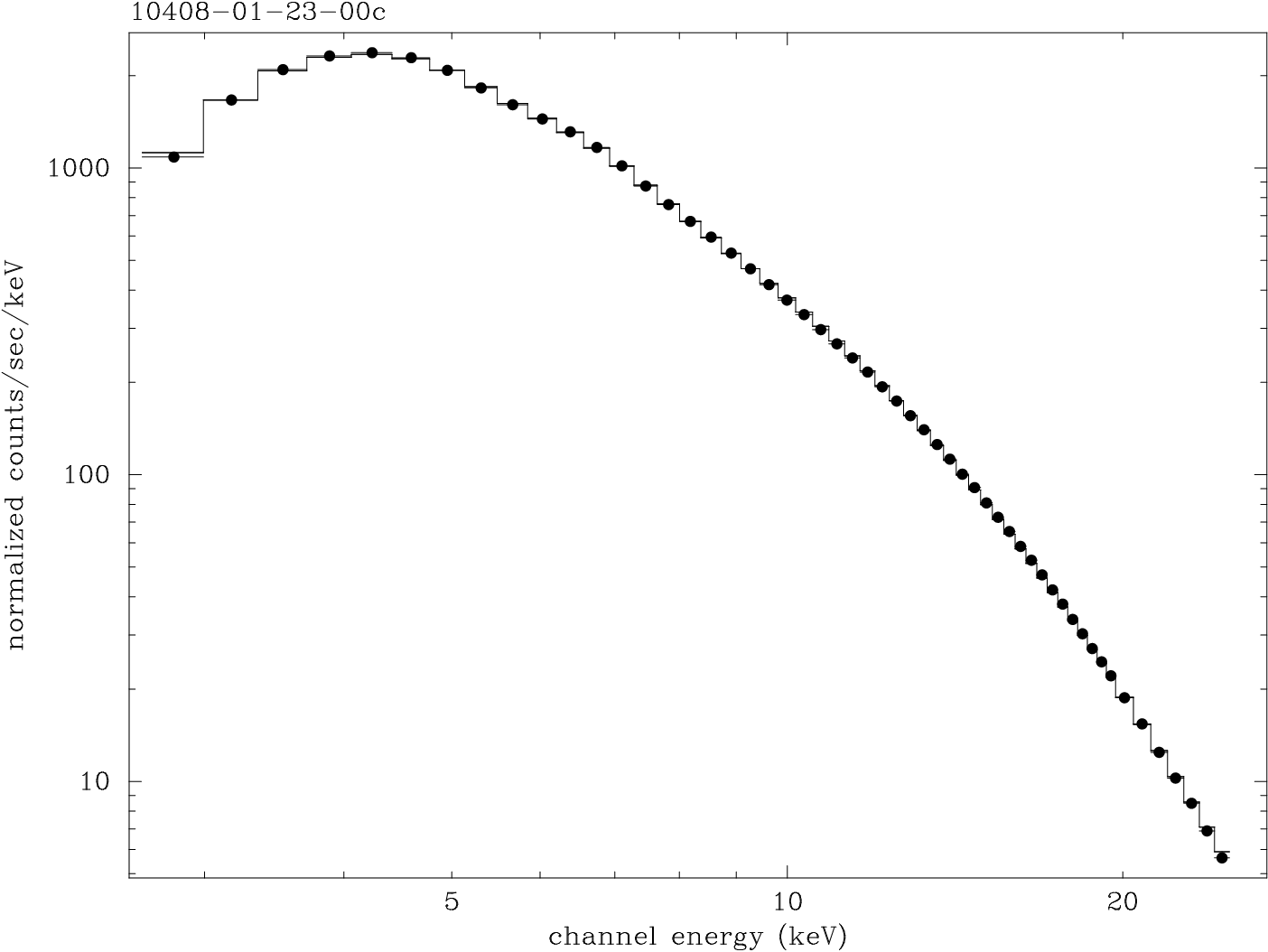,width=6cm,angle=-90,height=4cm}
\\
\end{tabular}
  \caption{Examples of power density spectra ({\it left panels}) and energy
spectra ({\it right panels}) for three plateau observations with different QPO 
frequency values.}
  \label{platqpo}
\end{figure*}

The plateau observations are relatively straightforward to analyze from the
point of view of timing analysis, because of the 
stable positive correlation between the QPO frequency and count rate
(see Reig et al. 2000). Therefore, no special timing selection
is necessary and it is possible to accumulate all the data to produce
one PDS and one energy spectrum.
To produce the PDS, for each observation we used
high-time resolution data in the 2-13 keV range. Some of the
observations have a time resolution of 1/512 s, limited to 1/128 s
for the others (see Table~\ref{tabobplat}). 
We divided each observation interval (orbit of the satellite) in intervals 16 
s long and produced an average PDS per 
observation summing the PDS of the individual segments. We subtracted the 
contribution due to Poissonian statistics and the Very Large Event window
(Zhang et al. 1995, Zhang 1995) from each PDS and renormalized them to squared
fractional rms per Hertz (see Belloni \& Hasinger 1990). Every PDS was then
logarithmically rebinned and fitted to find the centroid frequency of the 1-10
Hz QPO, which is very sharp 
and often shows higher harmonics and even a sub-harmonic component (as shown in
Fig.~\ref{platqpo} for three different observations). A model consisting of
the superposition of a number of Lorentzians (from 5 to 9 depending on the
observation) was used to fit every feature
present in the power spectrum (see Belloni et al.
2002). Because of the relative strength of the QPO
peak, the centroid frequency did not change appreciably when
using different models for the continuum noise: the QPO frequencies we found
with this analysis are in close agreement with those from Reig at
al. (2000). 
The resulting QPO centroid frequencies are listed in Table~\ref{tabobplat}.

In Class $\chi$ observations, the source is in a long steady state C (Belloni et
al. 2000). As shown by Belloni et al. (2000), this corresponds to an
undetectable thermal component in the PCA energy spectrum.
For each observation interval we extracted PCA energy spectra in the 
range 3--25~keV,
averaged over the whole interval, from spectral ({\tt Standard2}) data. 
For each spectrum,
we subtracted the background estimated with {\tt pcabackest v.~2.1e}, and
created an appropriate  detector response matrix with {\tt pcarsp v.~2.43}.
A systematic error of 1\% was added to the spectra to account for
remaining uncertainties in the detector calibration.
Our starting model for the spectral fits consisted of a power law, 
corrected for interstellar absorption,
plus a Gaussian emission line to take into account an excess at
6.4~keV. With this model, it was not possible to lower the reduced
$\chi^2$ below 2. We modified the model, adding a high-energy cutoff
to the power law. The resulting spectral parameters are shown in Table 1.
This model is consistent with that used by
Trudolyubov et al. (1999) and  Muno et al. (1999). Notice that Trudolyubov
et al. (1999) found rather high cutoff energies ($\sim$100 keV), while
Muno et al. (1999) found also much lower values, even below 10 keV.
Our values are in the range 20--30 keV. 
Only 2 among the selected plateau observations (10408-01-27-00 and
20402-01-52-00) were also analyzed by
Muno et al. (1999). They used an often-used model for Black Hole Candidates,
which consists of a superposition of a multicolor disk-blackbody (thermal
component) and a power law (without high energy cutoff). Their spectral
analysis results in a different power law photon index, $\sim 0.7$
grater than the values we found. However, they derive an extremely 
high inner disk
temperature ($\sim 4$~keV) and an inner disk radius consistent with zero
($\sim 2 \pm 2$ km). These parameters do not represent a realistic 
physical picture of the source
and therefore suggest that the inclusion of a disk-blackbody component
is not physically necessary.
\new{Our choice of a simple power law with a high-energy cutoff is motivated
by two reasons: the need of a simple model for efficiently performing a 
large number of fits, and the need to use a common model that allows
comparison with values in the literature. In order to check that
for high values of the spectral index $\Gamma$ this model yields meaningful
values, we fitted our spectra also with the {\tt bmc} model in XSPEC, which 
has been used by other authors to describe the energy spectra of Black
Hole Candidates in the framework of Comptonization models.
These fits give
values of $\Gamma$ consistent with those from a simple power-law model, with
a slight systematic excess of 3\%.}
\begin{figure}[h]
\centerline{\psfig{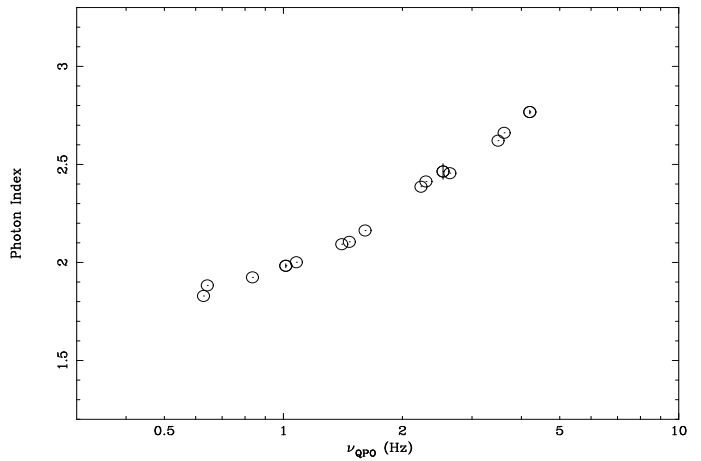}}
  \caption{Plot of Power Law Photon Index versus QPO centroid frequency for the
plateau observations of GRS~1915+105 studied. A positive correlation spanning
nearly an order of magnitude is evident. 
Representative error bars are plotted for three points.}
  \label{gammaqpoplat}
\end{figure}
\begin{figure*}[t]
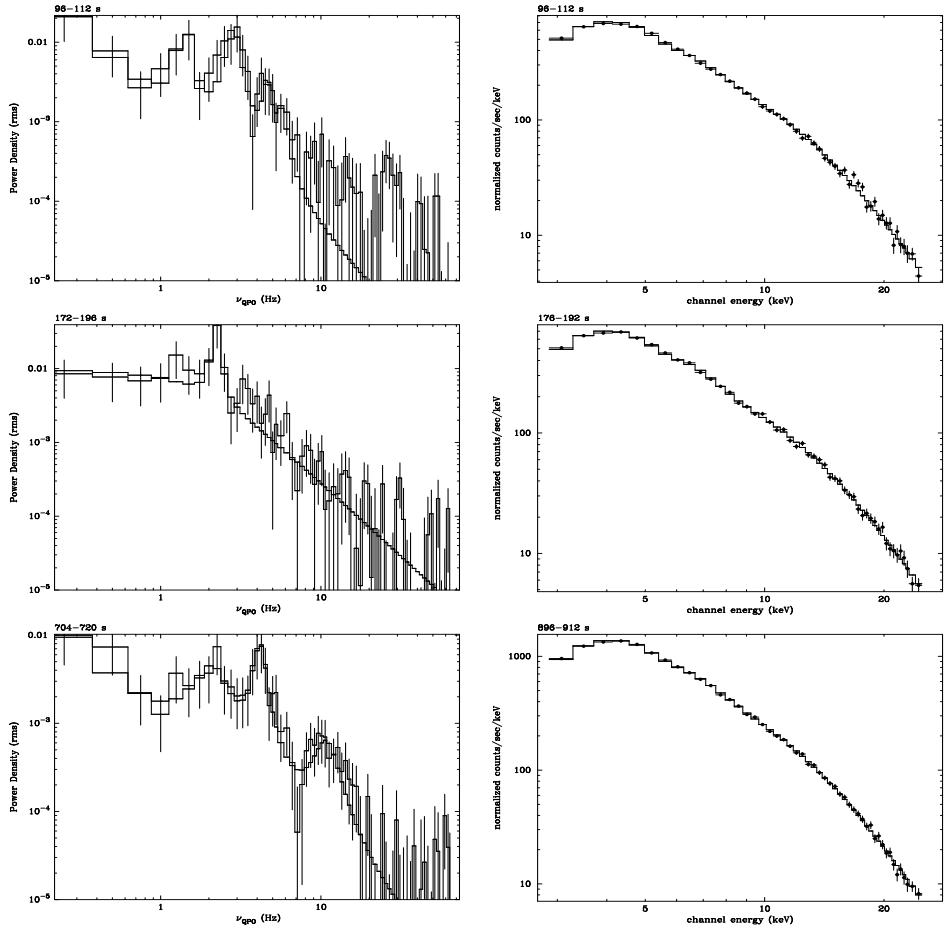

\centering
\begin{tabular}{cc}
\psfig{figure=h3835f3a.ps,width=6cm,angle=-90,height=4cm}
&
\psfig{figure=h3835f3b.ps,width=6cm,angle=-90,height=4cm}
\\
\psfig{figure=h3835f3c.ps,width=6cm,angle=-90,height=4cm}
&
\psfig{figure=h3835f3d.ps,width=6cm,angle=-90,height=4cm}
\\
\psfig{figure=h3835f3e.ps,width=6cm,angle=-90,height=4cm}
&
\psfig{figure=h3835f3f.ps,width=6cm,angle=-90,height=4cm}
\\
\end{tabular}
  \caption{Examples of power density spectra ({\it left panels}) and energy
spectra ({\it right panels}) for 40702-01-02-00 (class $\nu$) corresponding
to different time intervals. Start and end times for each panel are
from the observation start time (Table 2).}
  \label{otherqpo}
\end{figure*}
From Table~\ref{tabobplat}, one can see that there are variations in the
best fit values of the column density (${\rm N}_{\rm
H}$) (see also Belloni et al. 2000; Klein-Wolt et al. 2002).

In Fig.~\ref{gammaqpoplat} we plot the power-law photon index versus
the QPO centroid frequency. A positive correlation is evident. 
 
\subsection{Class $\alpha$, $\beta$ and $\nu$ observations}

\begin{figure}[!h]
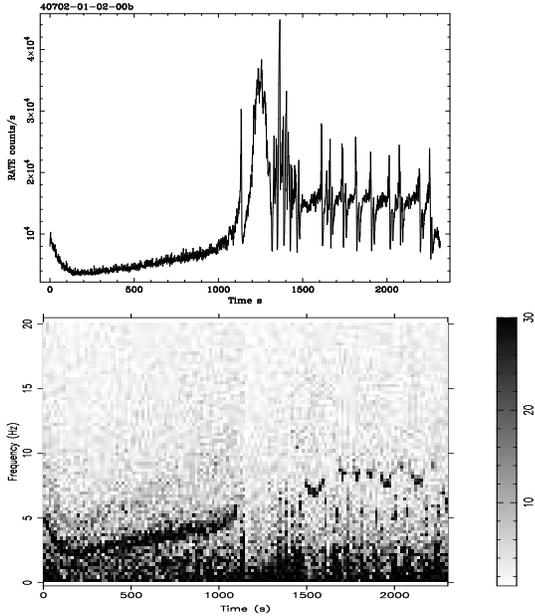

\centering
\begin{tabular}{l}
\psfig{figure=h3835f4a.ps,width=5.9cm,angle=-90,height=4cm}
\\
\psfig{figure=h3835f4b.ps,width=7cm,angle=-90,height=4cm}
\\
\end{tabular}
  \caption{Light curve and dynamical PDS for the $\nu$-class
observation 40702-01-02-00 (second interval). It is evident that
the 1-10 Hz QPO, when present, is strictly 
positively correlated with counts (see text).
In the PDS frame, frequency is on the y-axis while
the power scale is shown in the chromatic column on the right.}
  \label{fluxdyn}
\end{figure}

In order to examine the correlation between QPO centroid frequency and power-law
photon index, and in particular to extend it to higher values of the QPO
frequency, we also analyzed observations in
which GRS~1915+105 shows large variability on time scales longer than 1 second.
While belonging to different
classes, they all have in common the presence of at least one
state-C interval of more than 700 seconds (see Belloni et al. 2000).
The log of
selected observations is reported in Table~\ref{tabobother}.
After a close inspection of the light curves produced 
from {\tt Standard1} data (corresponding to the full PCA energy band 2--60 keV),
we identified the start and end times of the state-C intervals to analyze. 
Flux values during these periods
typically range from 3000 to 40000 cts/s. For each observation, we 
checked the choice of intervals 
by producing a dynamical PDS (a series of PDS as a function of time), 
with which we verified that a 1--10
Hz QPO was present throughout the whole interval. 
The dynamical PDS also
confirmed that the X-ray intensity of the source is strongly correlated 
with the QPO frequency (Chen et al. 1997, Markwardt et al. 1999, Reig et al. 
2000). Therefore,
following the PCA count rate, the QPO frequency first decreases rather fast,
then starts increasing at a lower rate (see Fig.~\ref{fluxdyn}).
\begin{figure}[h]
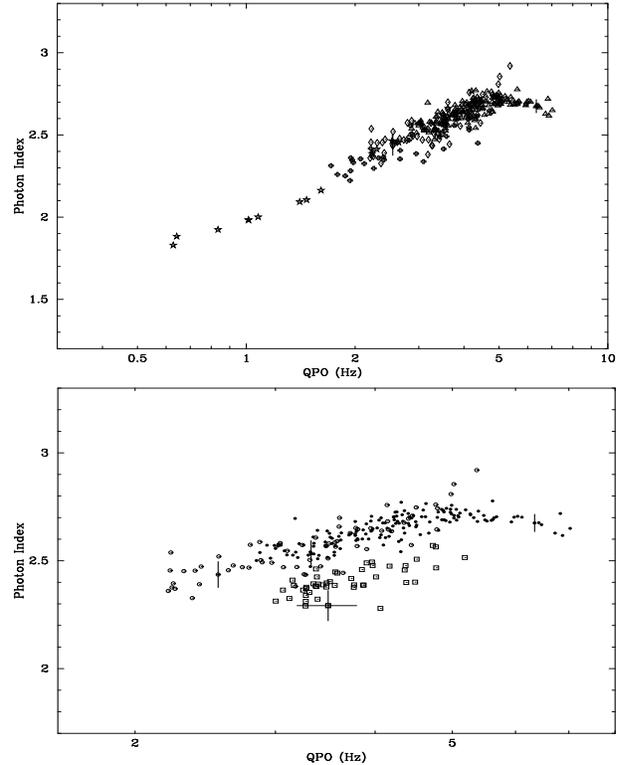

\centering
\begin{tabular}{c}
\psfig{figure=h3835f5a.ps,width=8cm,angle=-90,height=5cm}
\\
\psfig{figure=h3835f5b.ps,width=8cm,angle=-90,height=5cm}
\\
\end{tabular}
   \caption{{\it Upper panel}: power-law photon index vs. QPO
centroid frequency for the observations of class $\beta$ and $\nu$ studied. 
Values for the plateau observations (from Fig. 2)
are plotted for comparison (stars). Different symbols are used
to identify different observations. [Triangles= obs. 15,16 - Diamonds=
obs. 19 - Crosses= obs. 17].
{\it Lower panel}: power-law photon index vs. QPO centroid
frequency for the observations of class $\alpha$ and $\nu$ studied. 
Different symbols identify different observations. 
[Full circles= obs. 15,16 - Empty
circles= obs. 19 - Squares= obs. 18].
In both panels error bars at 90\% confidence
for three data points are shown as example.} 
   \label{res}
\end{figure}
\begin{table}[h]
   $$		
   \begin{array}{clccc}
    \hline
     
{\rm Obs. N }  &{\rm Obs.\, ID}   &{\rm MJD}  &{\rm Class}   &{\rm
\Delta T }  \\
     \hline
&&&&\\
15  & {\rm 10408-01-44-00}a & 50381.4942 & \nu     &1248  \\
16  & {\rm 10408-01-44-00}b & 50381.5655 & \nu     &1488  \\
17  & {\rm 20186-03-03-01}b & 50674.2347 & \beta   &704  \\
18  & {\rm 20187-02-01-00}b & 50575.9322 & \alpha  &928  \\
19  & {\rm 40702-01-02-00}b & 51320.9322 & \nu     &1120  \\

&&&&\\  
   \hline
   \end{array}
   $$
   \caption[]{List of the observations from classes $\alpha$, $\beta$
and $\nu$ analyzed in this work. 
Letters a, b, c correspond to different observation intervals within the same
observation. The last column refers to the length 
($\Delta$T) of the state-C interval
that we selected for our study} 
       \label{tabobother}
\end{table}

To perform timing analysis, we inspected 
the same type of data (high-time resolution and low energy range) as for
plateau observations. In this case however, 
we divided every state-C interval in segments with a length of 4 s
rather than 16, in order to better follow the rapid variations of the QPO
centroid frequency, and produced a PDS for every segment. After subtracting
the contribution from the Poissonian statistics and the 
Very Large Event window, we
summed these PDS in sets of four, rebinned them logarithmically and fitted the
result to find the QPO centroid frequency. 
As in the previous case, a model consisting of the superposition of a number
of Lorentzians was used (see Belloni et al. 2002).
We performed the summing process
because we wanted to correlate timing and spectral results (which have
a minimum time bin of 16 s), and this is
possible only if we analyze intervals with temporal coincidence.

Spectral parameters were obtained extracting energy spectra from {\tt
Standard2} data, with an  
integration time of 16~seconds, subtracting the background and creating the
detector response matrix as for the plateau observations. 
We fitted every spectrum
using the ``standard'' model for BHCs: the superposition of a multicolor
disk-blackbody and a power law for the high energy tail. We corrected for the
interstellar absorption, fixing ${\rm N}_{\rm H}=7\times 10^{22}$~cm$^{-2}$
(Klein-Wolt et al. 2002) and added a Gaussian emission line with central
energy fixed to 6.4~keV. 
With respect to the model adopted for the plateau observation, we needed
to include a disk-blackbody (see Belloni et al. 1997b, 2000) and we had to fix
${\rm N}_{\rm H}$ due to the limited statistics of the 16s spectra.
Again, we added in quadrature a systematic error of 1\%. 
The fits give acceptable results, with a
reduced $\chi^2$ usually around 1. Therefore, we can trace the evolution (with
a time step of 16~s) of the spectral parameters during the state-C intervals. We
can then compare the spectral and timing evolution by correlating the resulting
parameters.
\new{As in the case of plateau observations we fitted selected spectra with
the {\tt bmc} model (see Sect. 2.2), 
finding consistent results for $\Gamma$.}

Examples of both PDSs and energy spectra (with their best fit models)
can be seen in
Fig.~\ref{otherqpo}. The correlation between QPO centroid frequency and
power-law index for classes $\beta$ and $\nu$ is plotted
in Fig.~\ref{res} ({\it upper panel}). The behaviour is
similar to $\chi$ 
observations, although there is a spread due to the lower statistics available
from spectra averaged over only 16 seconds. At
high QPO frequencies it is possible to notice a ``turnoff'' in the correlation 
for observations 15 and 16, while observation 19 (diamonds) does not show this
feature.
Figure~\ref{res} ({\it lower panel}) shows the same correlation from classes 
$\alpha$ and $\nu$.
While $\nu$ data are compatible with those in the upper panel, $\alpha$ data 
show a correlation similar, but offset to lower photon-index values.

\section{Other sources}

In order to compare our results for GRS 1915+105
with other systems, we searched the literature for similar data from 
other black hole candidates. We identified four systems for which
values for the QPO centroid frequency and the power-law index are
available, and for which there are at least a few different values.
Notice that, as mentioned above, we do not consider parameters like
inner disk radius, which are difficult to compare between different
sources. Providing the adopted spectral models are compatible, the values
of $\Gamma$ are directly comparable, and so of course are the QPO centroid
frequencies.

\subsection{GRO J1655-40}

This source is a BHC with a dynamical estimate for the mass of the 
black hole of $\sim
6.3  {\rm M}_{\odot}$ (Bailyn et al. 1995; Greene et al. 2001).
The system, a bright X-ray transient which appeared for the first time in 1994
and showed a subsequent outburst in 1996/1997, showed
superluminal radio jets (Hjellming \& Rupen 1995). GRO J1655-40 is therefore
another of the few objects belonging
to the {\it microquasar} class. This object shows 
quite a few different QPO peaks in a range between 0.1
and 450 Hz, (see Remillard et al. 1999a;
Strohmayer 2001, Remillard et al. 2002b).
Among them, we are interested in the variable 14--22~Hz QPO (see Sobczak et
al. 2000a).
We consider here PCA/RXTE data from an intensive observational campaign
of the source during its second outburst in 1996--1997. The timing parameters
(i.e., the QPO centroid frequencies) are taken from Table 3 of Remillard et
al. (1999a), and the spectral parameters (i.e., the power law photon 
indexes) from
Table 1 of Sobczak et al. (1999a). The energy spectra in the 2.5--20~keV energy
range were fitted using a model consisting of a multicolor disk-blackbody
plus a power law; a detailed description of the spectral analysis can be
found in Sobczak et al. (1999a). The power density spectrum for each observation
was analyzed in the 2--30~keV energy range, obtaining the QPO central frequency
fitting with a Lorentzian function plus a power law for the local 
continuum (see Remillard et al. 1999a).  
The plot of power-law index versus QPO centroid frequency for GRO J1655-40
is shown in Fig.~\ref{other1} (squares). 
There is a clear anti-correlation between
these two parameters, as already shown by Sobczak et al. (1999a).

\subsection{XTE J1550-564}

This BHC is the brightest X-ray transient observed
with RXTE (the 1998--1999 flare reached 6.8~Crab in the 2--10~keV band). In the
PDS, it shows both low-frequency (0.8--18 Hz) and high-frequency (180--280 Hz)
QPOs (Cui et al. 1999; Remillard et al. 1999b; Homan
et al. 2001; Remillard et al. 2002a,b). 
We consider here the low-frequency one (see Remillard
et al. 2002a).
The observations presented here are the whole set made with PCA/RXTE 
before 1999
March 22 (Gain Epoch 3). The power-law photon indexes obtained from the  
spectral analysis of these observations are taken from Table 1 of Sobczak et
al. (1999b), while the QPO centroid frequencies from the PDS analysis are listed
in Table 1 of Sobczak et al. (2000a). The spectral model used by Sobczak et 
al. (1999b) to fit the
2.5--20~keV energy spectra consists again of the superposition of a multicolor
disk-blackbody and a power law, with corrections for interstellar absorption
and a Gaussian emission line at $\sim 6-7$~keV. 
The centroid frequency of the QPO 
was found fitting the peak with a Lorentzian function and the continuum with a
power law.

\begin{figure}[h]
\centerline{\psfig{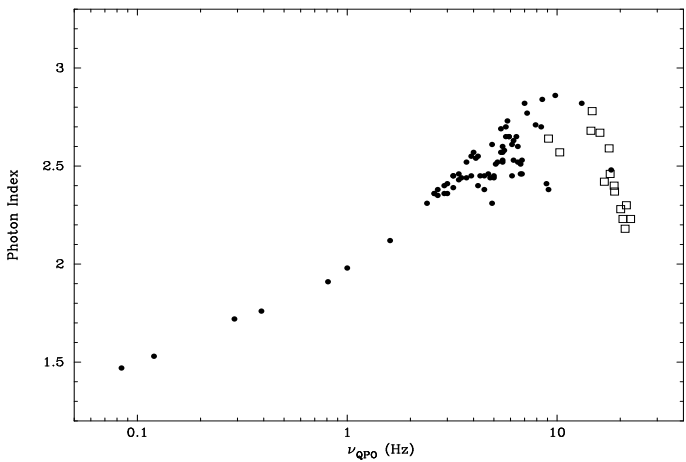}}
  \caption{Power-law photon index vs. QPO centroid frequency for
{\it XTE J1550-564} (circles) and {\it GRO J1655-40} (squares). 
While for the first source there
is a general positive correlation between the two parameters, 
for the second one the correlation is negative. However, one can
see that data points of {\it
GRO J1655-40} seem to extend the ``turnoff'' curves of
{\it XTE J1550-564} (see text).} 
  \label{other1}
\end{figure}
Data obtained from the  literature for XTE J1550-564 are plotted in
Fig.~\ref{other1} (circles). They show a positive correlation 
between power law photon index and QPO centroid frequency very similar to the
behaviour of GRS~1915+105 (see Fig. \ref{gammaqpoplat} and \ref{res}). 
However, while the 
correlation is very good at low frequency, at higher QPO values there
is a larger spread. 
Surprisingly, the points from GRO J1655-40, although showing a 
negative correlation, are placed on the natural
continuation of the upper points for XTE J1550-564.
 As one can see from Cui et al. (1999), Remillard et al. (2002b) and Homan
et al. (2001), the low-frequency QPO in XTE J1550-564 does not maintain the 
same properties throughout the whole outburst. It is therefore natural to 
examine its behaviour in separate intervals of the light curve. 

We subdivide
the light curve in four separate sections (see Fig. 7): a `rise' interval
from the first observation up to the peak flux, a `down' interval from the 
peak following the smooth decrease in flux until the first small flare,
a `flat' interval where the flux remains rather stable, with small flares,
and a `decay' interval where the flux decays at the end of the first part
of the outburst. After the last observation examined here, no QPO was
detected for a long time, until the flux rose again to the second
part of the outburst (see Homan et al. 2001). 

In Fig. 8 the behaviour of
XTE J1550-564 in the QPO-$\Gamma$ plane is shown with different symbols for
the four different intervals. One can see that the first two intervals
trace a very smooth curve in Fig. 8, the second interval tracing back the
path of the first. The scattered points correspond to the remaining two
intervals, and it is interesting to note that all these points are located
below the others. What is different between the first two and the second 
two classes of observations
that have a different behaviour in this plot? From Sobczak et al. (2000b)
one can see that a clear dividing line between the two groups is the
ratio of the power-law flux over the total flux, estimated from the observed
PCA spectra. The first observations have a lower value of the ratio. Also
notice that the points that contribute to the scatter in Fig. 8 are the ones
from the `decay' interval, which contains QPOs classified as A/B in
Homan et al. 2001, while the QPO observed
in the previous intervals is classified as C (notice that these QPO classes
have nothing to do with the A/B/C classes defined in Belloni et al. 2000).

\begin{figure}[h]
\centerline{\psfig{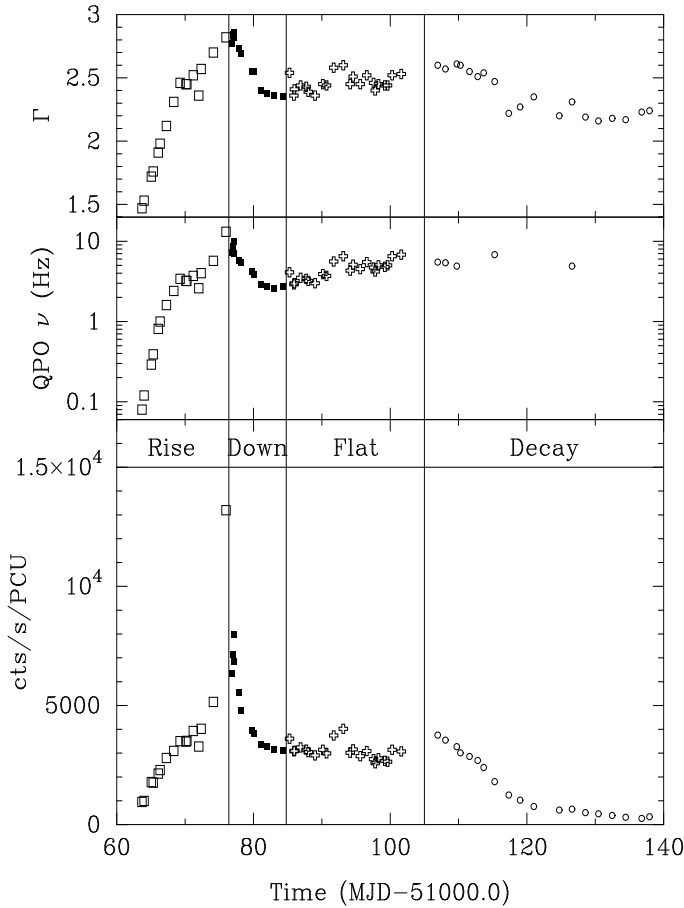}}
  \caption{Bottom panel: light curve for 
XTE J1550-564 as derived from Table 1 of
Sobczak et al. (2000b). Observations span from 7th September 1998 to 20th
November 1998. Different symbols correspond to the different intervals
marked in the panel (see also Fig. 8). Middle panel: corresponding QPO
frequencies (from Sobczak et al. 2000a). Top panel: corresponding power-law
slopes (from Sobczak et al. 1999b).}
  \label{curva15}
\end{figure}

\begin{figure}[h]
\centerline{\psfig{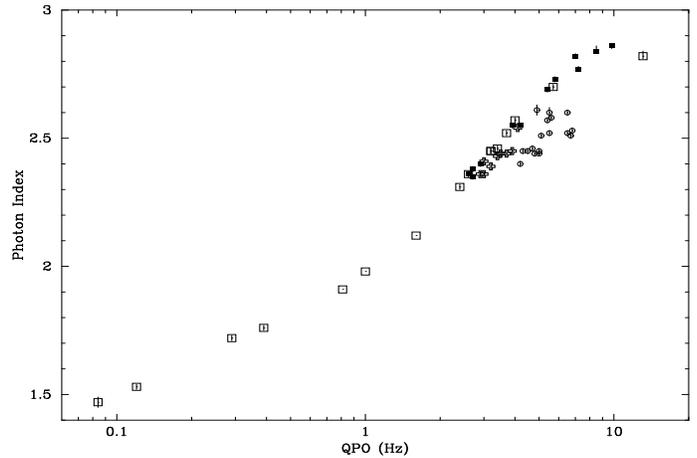}}
  \caption{Plot of Power Law Photon Index versus QPO centroid frequency for
{\it XTE J1550-564}. Data points are labeled with different marks according
to the selection explained in text (same symbols as in Fig. 7).}
  \label{quindici}
\end{figure}

\subsection{XTE J1748-288}

This BHC shows superluminal radio jets and belongs to
the class of galactic microquasars, for the same reasons as GRS~1915+105 and GRO
J1655-40 (Smith et al. 1998; Hjellming et al. 1998;
Revnivtsev et al. 2000; Kotani et al. 2000;
Miller et al. 2001). 
During the 1998 outburst the source was observed to display spectral and
timing properties typical for BHCs systems (Revnivtsev, et al. 2000). 
A QPO feature
was found in the PDS at 0.5~Hz and 20--30~Hz. Here we consider spectral
and timing parameters given in Table 2 and 3 of Revnivtsev et al. (2000) who
reported an analysis of the PCA/RXTE observations during the 1998
outburst. They fitted the energy spectra between 3 and 25~keV using the
standard model for BHCs consisting of a multicolor disk-blackbody plus a power 
law, with low energy interstellar absorption and a Gaussian emission line at
$\sim 6.5$~keV. For the seven observations considered here, Revnivtsev et
al. (2000) fitted the PDS using the sum of a flat-topped band limited
noise component, a power law component, and a Lorentzian function to describe
the QPOs. The resulting points in the QPO-$\Gamma$ plot are shown as
empty diamonds in Fig. 9. There is a clear positive 
correlation between $\Gamma$ and QPO, but it is not identical to the same
correlations of the three sources previously discussed. Notice that the
QPO frequencies are the highest considered here.

\subsection{4U 1630-47}

This X-ray transient is considered a BHC for which
the optical thin radio emission, observed during the 1998 outburst, suggests
the presence of radio lobes (Hjellming et al. 1999). 
We obtained timing (QPO centroid frequency) and spectral 
(power-law index) parameters
of the observations made during the
decay of the 1998 outburst from Table 4 and 5 of Tomsick \& Kaaret
(2000). The timing analysis was made in the 2--21~keV energy band fitting
the PDS with a flat-top component for the continuum and a Lorentzian function 
for the QPO feature. The centroid frequency was between 0.2 and 3.4 Hz. 
The energy spectra were fitted in the 2.5--20~keV range using a
power law, a multicolor disk-blackbody soft component, a Gaussian
emission line at $\sim 7$~keV, the interstellar absorption and a broad iron
absorption edge.
For the peak of the outburst, we used the results of Trudolyubov et al. 
(2001). Here timing analysis was performed from data in the 2--13 keV
energy range and the PDS were fitted with Lorentzian models plus a
low frequency power-law. The PDS were rather complex: as centroid frequency
we used the peak at the lowest frequency in the 1--20 Hz range.
The 3-20 keV PCA energy spectra were fitted with a disk-blackbody plus a 
power law, although the presence of an excess around 6--8 keV was recognized.

\begin{figure}[h]
\centerline{\psfig{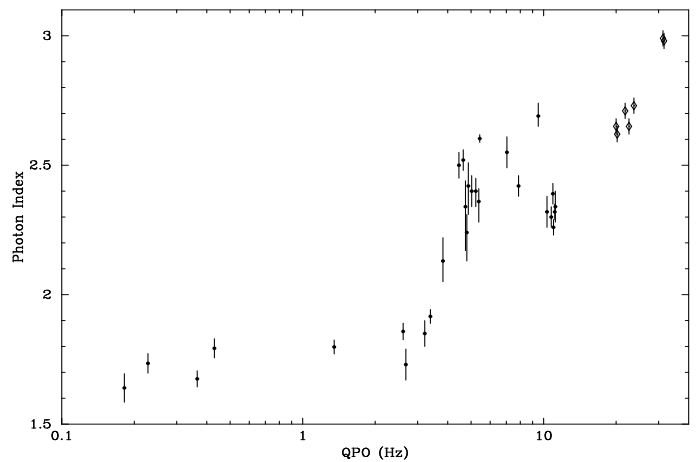}}
  \caption{Plot of Power Law Photon Index versus QPO centroid frequency for
{\it XTE J1748-288} (diamonds) and {\it 4U 1630-47} (filled
circles).}
  \label{other2}
\end{figure}
The points corresponding to this source are shown in Fig.~\ref{other2}. 
It is possible to recognize a behaviour similar to those shown
by GRS~1915+105 and XTE~J1550-564, although in this case the correlation is
not as smooth. The points corresponding to the end phase of the outburst
have a rather constant $\Gamma$ around 1.7--1.8, while the QPO frequency
decreases from 3 Hz down to 0.2 Hz (data from Tomsick \& Kaaret 2000). 
Notice that the points corresponding
to the lowest frequencies (0.2 to 0.5 Hz) do match rather well the 
correlation followed by XTE J1550-564. At higher QPO frequencies
there is a correlation between the two quantities, with a rather
large scatter and a possible indication of a turn-off.

\begin{figure*}[!b]
\centerline{\psfig{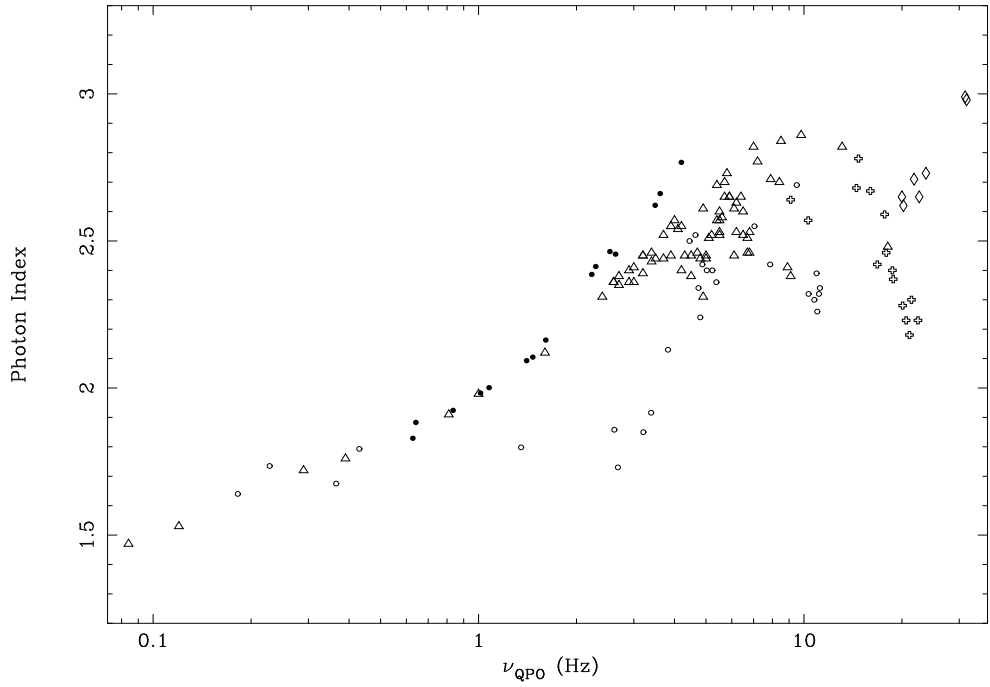}}
  \caption{Plot of Power Law Photon Index versus QPO centroid
frequency. Plateau results from GRS~1915+105 are plotted in full circles;
empty crosses are data from GRO J1655-40; empty circles are data from
4U~1630-47; diamonds are data from XTE J1748-288; empty triangles are data
from XTE J1550-564.}
  \label{final}
\end{figure*}

\section{Summary and discussion}

Our results can be summarized as follows. For the $\chi$ class of GRS 1915+105,
when the energy spectrum is rather hard and consists exclusively of 
the power-law component and the power density spectrum shows a clear
1--10 Hz QPO, there is a strong positive correlation between QPO centroid
frequency and photon index of the power law. At the highest values of the
QPO centroid frequencies, this correlation flattens and appears to start
reversing. Published data from other sources indicate that this general
correlation exists for a number of sources, and indeed in some cases there
is a turnoff in the relation for high QPO frequencies. The correlation 
is not exactly the same for all sources and in the case of XTE J1550-564
there is evidence for different turnoffs during different intervals of the
source outburst.

The turnoff in the correlation is difficult to understand. If the QPO
frequency is associated to a specific radius in the accretion disk,
only the rising part of the correlation can be explained with a
simple Comptonization model. If the QPO-radius relation also shows
a reversal (see for instance Rodriguez et al. 2002 and 
Varni\'ere et al. 2002), then a turnoff like that presented
here can be produced.

The source for which we have more data, and the largest range in QPO
frequencies is XTE J1550-564. From Fig. 8 one can see that not only
the points from the first two sections of the outburst (see Fig. 7) 
follow a very narrow correlation, but also that all subsequent points 
lie below this correlation. This suggests that the correlation followed
in the first part, when the overall PDS was in the form of a flat-top
noise component plus a QPO (Cui et al. 1999; Remillard et al. 2002b), 
is perturbed in a very specific way when 
the source changes state and a powerful disk component appears in the
energy spectrum (Sobczak et al. 1999b,2000a,b).
There is increasing
evidence that the power-law component observed in the energy spectra of
BHCs in the Low/Hard state is physically different from that seen in the
Very High State (see e.g. Zdziarski et al. 2001). It is therefore possible
that the turnoff is due to a gradual transition between these two regimes.

\new{It is interesting}
to discuss the position of the turnoff in the QPO-$\Gamma$
relation. Figure 10, where all the sources are shown together,  
suggests that: (a) XTE J1550-564 and GRO J1655-40 have a similar turnoff
frequency around 10 Hz; (b) XTE J1748-288 has a higher turnoff frequency, 
if any, larger than 30 Hz; (c) GRS 1915+105 (see Fig. 5) has a lower turnoff
frequency at about 5 Hz.
It is tempting to compare these values with the relative values of the
estimated dynamical masses for the compact objects in the systems. Excluding
XTE J1748-288, for which no dynamical mass estimate is available, we see
that GRS 1915+105 has a mass roughly twice and
a turnoff frequency half that of GRO J1655-40, as one could expect 
with a simple mass scaling. However, the recent mass estimate for 
XTE J1550-564 is around 10$_\odot$ (Orosz et al. 2002), from which we
would expect a turnoff frequency lower by a factor of $\sim$2 with
respect to that of GRO J1655-40, which is evidently not observed.

In conclusion, we have found that in GRS 1915+105 and a number of other
BHCs, the centroid frequency of the low-frequency QPO and the photon index
of the power-law component in the energy spectra follow a very similar
kind of correlation. In particular, for some sources there is evidence for the
presence of a turnoff in this correlation. If the QPO frequency is 
associated to a specific radius of the accretion disk, this provides an
important clue to the possible origin of the power-law component.
More data are needed in order to understand the physical nature of the
turnoff, but these results stress the importance of such simple
spectral/timing correlations for black hole candidates.

\begin{acknowledgements}
TB acknowledges the hospitality of MIT and thanks the Cariplo Foundation for 
financial support.

\end{acknowledgements}

\end{document}